\documentclass{JHEP3}
\usepackage{latexsym}
\usepackage{epsfig,amssymb,euscript}
\usepackage{amsmath}
\usepackage{array,calc,epsfig}
\usepackage{citesort}
\def\be{\begin{equation}}
\def\ee{\end{equation}}
\def\bea{\begin{eqnarray}}
\def\eea{\end{eqnarray}}
\newcommand{\ul}{\underline}
\numberwithin{equation}{section} %%
\usepackage[]{graphicx}

\def\calf         {{\cal F}}

\def\tr           {\mathop{\rm Tr}}
\def\Re           {{\rm Re\hskip0.1em}}
\def\Im           {{\rm Im\hskip0.1em}}

\def\sqr#1#2{{\vcenter{\vbox{\hrule height.#2pt
 \hbox{\vrule width.#2pt height#1pt \kern#1pt \vrule width.#2pt}\hrule
 height.#2pt}}}}

\def\a{\alpha}

\def\slashchar#1{\setbox0=\hbox{$#1$}           % set a box for #1
\dimen0=\wd0                                 % and get its size
\setbox1=\hbox{/} \dimen1=\wd1               % get siste of /
\ifdim\dimen0>\dimen1                        % #1 is bigger
\rlap{\hbox to \dimen0{\hfil/\hfil}}      % so center / in box
#1                                        % and print #1
\else                                        % / is bigger
\rlap{\hbox to \dimen1{\hfil$#1$\hfil}}   % so center #1
/                                         % and print /
\fi}

\title{Supersymmetric D-branes and calibrations on general ${\cal N}=1$ backgrounds}

\author{Luca Martucci$^a$ and Paul Smyth$^{ab}$\\ $^a$Institute for Theoretical Physics, K.U. Leuven,\\ $~$Celestijnenlaan 200D, B-3001 Leuven, Belgium \\ $^b$The Blackett Laboratory, Imperial College London, \\$~$Prince Consort Road, London SW7 2AZ, U.K. \\$~$Email: \email{luca.martucci@fys.kuleuven.be, paul.smyth@imperial.ac.uk}}

\abstract{We study  the conditions to have
supersymmetric  D-branes on general ${\cal N}=1$ backgrounds with
Ramond-Ramond fluxes. These conditions can be written in terms of
the two pure spinors associated to the $SU(3)\times SU(3)$
structure on $T_M\oplus T^\star_M$, and can be split into two
parts each involving a different pure spinor. The first involves
the integrable pure spinor and requires the D-brane to wrap a
generalised complex submanifold with respect to the generalised
complex structure associated to it. The second contains the
non-integrable pure spinor and is related to the stability of the
brane. The two conditions can be rephrased as a generalised
calibration condition for the brane. The results preserve the
generalised mirror symmetry relating the type IIA and IIB
backgrounds considered, giving further evidence for this duality.}


\preprint{KUL-TF-05/16 \\Imperial/TP/050701 \\hep-th/0507099}

\begin{document}

\section{Introduction}

The study of string and supergravity backgrounds with fluxes has received much attention in the recent years  due to the key role that they play in many interesting situations. For example, they appear to be fundamental in the search for more realistic and phenomenologically interesting stringy scenarios, and also in the construction of string models holographically dual to relevant gauge theories. As fundamental objects of the theory, D-branes occupy a preeminent position in all these models and several aspects of their physics in such nontrivial situations deserve a better understanding.

In this paper we aim to study the geometry of supersymmetric
D-brane  configurations in a very general class of supergravity
backgrounds preserving four-dimensional Poincar\'e invariance and
${\cal N}=1$ supersymmetry. Such backgrounds correspond to warped
products of the four-dimensional Minkowski space-time and an
internal six-dimensional manifold $M$ with general fluxes turned
on. ${\cal N}=1$ supersymmetry requires the existence of four
independent 10d killing spinors, whose most general form can be
written in terms of two internal six-dimensional Weyl spinors
$\eta^{(1)}_+$ and $\eta^{(2)}_+$. This implies that $M$  has a
reduced $SU(3)\times SU(3)$ structure on $T_M\oplus T^\star_M$,
which may be further restricted to a $SU(3)$ or $SU(2)$ structure
on $T_M$. As discussed in \cite{gmpt,gmpt2}, these vacua can be
nicely characterised in terms of two $O(6,6)$ {\em pure spinors}
$\slashchar\Psi^\pm=\eta^{(1)}_+\otimes \eta^{(2)}_\pm$. Using the
Clifford map, the pure spinors  can be equivalently seen as formal
sums of forms $\Psi^\pm=\sum_k\Psi^\pm_{(k)}$, where $k$ is even
for $\Psi^+$ and odd for $\Psi^-$.

This formalism introduces a natural relation to generalised
complex geometry \cite{hitchin,gualtieri}\footnote{See
\cite{Witt} for previous discussions on the use of $SU(3)\times
SU(3)$ and other ``generalised" structures to describe
supersymmetric type II compactifications in the context of
generalised geometries.}. The two pure spinors are associated to
{\em generalised almost complex structures} whose (generalised)
integrability corresponds in turn to `closureness' of the pure
spinors under the twisted derivative operator $d_H=d+H\wedge$. In
\cite{gmpt2} it has been shown that the supersymmetry conditions
provide the integrability of the almost complex structure
associated to one pure spinor and that it defines a twisted
generalised  Calabi-Yau (CY) structure \`a la Hitchin
\cite{hitchin} on the internal manifold. On the other hand,  the
second pure spinor is not integrable due to the presence of
Ramond-Ramond (RR) field-strengths which act as  an obstruction to
integrability. As a consequence, if for example we restrict
ourselves to the $SU(3)$ case, the internal manifold will be
either symplectic (IIA) or complex (IIB). In the more general
$SU(3)\times SU(3)$ case, the manifold is a complex-symplectic
hybrid, even if IIA and IIB continue to ``prefer'' symplectic and
complex manifolds respectively \cite{gmpt2}.

In the following sections we will see how it is possible to
characterise  the supersymmetric D-brane configurations completely
in terms of the two pure spinors for this general class of ${\cal
N}=1$ backgrounds. We will mainly focus on the case of 
branes filling the flat 4d space-time  and the resulting 
equations [see equations (\ref{gens1}) and (\ref{gens2}) or 
equivalently (\ref{cc1}) and
(\ref{cc2})] represent the generalisation to ${\cal N}=1$ flux
backgrounds of the conditions obtained in \cite{bbs,mmms} for
branes wrapped on cycles of $CY_3$. This can be seen from the form
these conditions take once we restrict to the $SU(3)$ case [see
equations (\ref{minksusy1}) and (\ref{minksusy2}) or
equivalently (\ref{cc3}) and (\ref{cc4})], which can be considered
as formally the closest to the CY case\footnote{Equivalent
conditions have recently been presented for D-branes on IIB
$SU(3)$-structure backgrounds in \cite{marchesano}, where several
interesting applications to the warped Calabi-Yau subcase
\cite{gp} are also discussed.}. One preliminary necessary
requirement in order to get supersymmetric branes is that the two
internal spinors $\eta^{(1)}$ and $\eta^{(2)}$ must have the same
norm. Then,  the supersymmetry conditions split into two parts
involving the two pure spinors $\Psi^\pm$ and are completely
symmetric under the exchange $\Psi^+ \leftrightarrow \Psi^-$  as
one goes from type IIA backgrounds to type IIB and vice-versa.
This symmetry can be seen as a generalisation of the usual mirror
symmetry between supersymmetric cycles on standard CY's.

The first supersymmetry condition for a space-time filling D-brane wrapping an
internal $n$-cycle can be written in the form \bea\label{a1}
\Big\{ P[(g^{mk}\imath_k+dx^m\wedge)\Psi]\wedge
e^\calf\Big\}_{(n)}=0 \ , \eea where $\calf=f+P[B]$ ($f$ is the
world-volume field-strength), $\Psi$ is equal to $\Psi^-$ in IIB
and $\Psi^+$ in IIA, $P[.]$ indicates the pullback on the
worldvolume of the brane, and in the left hand side we consider
only forms of rank equal to the dimension $n$ of the wrapped
cycle. These pure spinors are exactly the integrable ones for each
case and we will discuss how this condition means that
supersymmetric cycles are generalised complex submanifolds with
respect to the appropriate integrable generalised complex
structure $\cal J$, as defined in \cite{gualtieri}. Then,
supersymmetric branes wrap an appropriate generalisation of a
complex submanifold in type IIB and of coisotropic submanifolds in
type IIA, and this identification becomes precise in the
$SU(3)$-structure case. This result is completely analogous to the
one recently discussed in \cite{koerber} where  D-branes on
supersymmetric backgrounds with only nontrivial Neveu-Schwarz (NS)
fields are considered (for previous work on branes in the context
of generalised complex geometry see
\cite{Kapustin,zabzine,Zucchini,Li,Kapustin:2005vs}).

The second supersymmetry condition is related to the stability of
the D-brane and can be written as \bea\label{a2}
\Big\{\Im\big(iP[\Psi]\big)\wedge e^\calf\Big\}_{(n)}=0\ , \eea
where now $\Psi$ is equal to $\Psi^+$ in IIB and $\Psi^-$ in IIA
(i.e. is the non-integrable pure spinor). The two conditions
(\ref{a1}) and (\ref{a2}) imply that for a suitable choice of
orientation on the wrapped cycle, the D-brane configuration is
supersymmetric. Since we are considering backgrounds with
nontrivial RR fluxes turned on, reversing the orientation on the
brane does not generally preserve  supersymmetry.

The above two conditions can be rephrased in terms of a  single
condition which also encodes the necessary orientation
requirement. For a D-brane wrapping an internal $n$-cycle, this is
given by \bea \Big\{\Re\big(-iP[\Psi]\big)\wedge
e^\calf\Big\}_{(n)}=\frac{||\Psi||}{8}\sqrt{-\det(g+\calf)}d\sigma^1\wedge\ldots
d\sigma^n\ , \eea where again $\Psi$ is equal to $\Psi^+$ in IIB
and $\Psi^-$ in IIA and
$||\Psi||^2=\tr(\slashchar\Psi\slashchar\Psi^\dagger) $. This
condition will be identified as a calibration condition with
respect to an appropriate {\em generalised calibration}
$\omega=\sum_k\omega_{(k)}$, with $\omega_{(k)}$ being a $k$ form,
which by definition is twisted closed, i.e. $d_H\omega=0$, and
must fulfil a condition of minimisation of the D-brane energy
density. More specifically, for any space-time filling D-brane
wrapping any internal cycle $\Sigma$ and with any worldvolume
field strength $\calf$ (such that $d\calf=P_\Sigma[H]$), we must
have \bea P_\Sigma[\omega]\wedge e^\calf\leq {\cal
E}(\Sigma,\calf)\ , \eea where ${\cal E}$ represents the energy
density [see equation (\ref{endens})] and in the left hand side we
mean that only forms of rank equal to the dimension of the wrapped
cycle are considered. An analogous definition of generalised
calibration has recently been used in \cite{koerber} for the case
with only nontrivial NS fields, and our result represents an
extension of that proposal in presence of non-zero RR fluxes.

The paper is organised as follows. In section 2 we review the basic conditions defining the general class of ${\cal N}=1$ backgrounds we are considering. In section 3 we derive the supersymmetry conditions for supersymmetric D-branes using $\kappa$-symmetry and express them in terms of the pure spinors $\Psi^\pm$ characterising our backgrounds. In section 4 and 5 we clarify the meaning of the conditions for the internal supersymmetric cycles, identifying them as generalised complex submanifold calibrated with respect to the appropriate  definition of generalised calibration. Finally we present our conclusions. Appendix A contains some basic properties of the almost complex structure and (3,0)-form that can be constructed from an internal spinor. Appendix B presents some details on the calculation of the background supersymmetry conditions needed in our analysis.

%%%%%%%%%%%%%%%%%%%%%%%%%%%%%%%%%%%%%%%%%
\section{Basic results on ${\cal N}=1$ vacua}\label{sec1}
We are interested in type II warped backgrounds preserving four-dimensional Poincar\'e invariance and ${\cal{N}}=1$ supersymmetry, with the most general fluxes and fields turned on. The ansatz for the ten dimensional metric $G_{MN}$ is
\bea
ds^2=e^{2A(y)}dx^\mu dx_\mu+ g_{mn}(y)dy^m dy^n\ ,
\eea
where $x^\mu$, $\mu=0,\ldots,3$ label the four-dimensional flat space, and $y^m$, $m=1,\ldots,6,$ the internal space.
Let us introduce the modified RR field strengths
\bea
F_{(n+1)}=dC_{(n)}+H\wedge C_{(n-2)}\ ,
\eea
where $dC_{(n)}$ are the standard RR field strengths\footnote{We will essentially follow the conventions of \cite{gmpt,gmpt2}, up to some differences consisting  in a sign for $H$ in type IIB  and the sign change
$C_{(2n+1)}\rightarrow (-)^nC_{(2n+1)}$ in type IIA.}.
In order to preserve four dimensional Poincar\'e invariance we can write
\bea\label{RRdec}
 F_{(n)}=\hat F_{(n)}+Vol_{(4)}\wedge \tilde F_{(n-4)}\ .
\eea The relation $F_{(n)}=(-)^{\frac{(n-1)(n-2)}{2}}\star_{10}
F_{(10-n)}$ between the lower and higher rank  field strengths
translates into a relation of the form $\tilde
F_{(n)}=(-)^{\frac{(n-1)(n-2)}{2}}\star_6 \hat F_{(6-n)}$ between
their internal components. The ten dimensional gamma matrices
$\Gamma_{\ul{M}}$ (underlined indices correspond to flat indices)
can be chosen in a real representation and  decomposed in the
following way \bea \Gamma_{\ul\mu}=\gamma_{\ul\mu}\otimes 1
\quad,\quad \Gamma_{\ul m}=\gamma_{(4)}\otimes \hat\gamma_{\ul m}\
, \eea where the four-dimensional gammas $\gamma_{\ul\mu}$ are real and the six-dimensional
ones $\hat\gamma_{\ul m}$ are anti-symmetric and purely imaginary.
The four- and six-dimensional chirality operators are given respectively by \bea
\gamma_{(4)}=i\gamma^{\ul{0123}}\quad,\quad
\hat\gamma_{(6)}=-i\hat\gamma^{\ul{123456}}\ , \eea so that the
10d chirality operator can be written as
$\Gamma_{(10)}=\Gamma^{\ul{0\cdots
9}}=\gamma_{(4)}\otimes\hat\gamma_{(6)}$.

For type IIA backgrounds the supersymmetry parameter is a 10d
Majorana spinor $\varepsilon$ that can be split in two
Majorana-Weyl (MW) spinors of opposite chirality: \bea
\varepsilon=\varepsilon_1 +\varepsilon_2\quad,\quad
\Gamma_{(10)}\varepsilon_1=\varepsilon_1\quad,\quad
\Gamma_{(10)}\varepsilon_2=-\varepsilon_2\ . \eea Since we are
interested only in four-dimensional ${\cal N}=1$ backgrounds, they
must have 4 independent Killing spinors that can be decomposed as
\bea\label{spinors} \varepsilon_1(y) &=&\zeta_+\otimes
\eta^{(1)}_{+}(y)+\zeta_-\otimes \eta^{(1)}_{-}(y)\ ,\cr
\varepsilon_2(y) &=&\zeta_+\otimes
\eta^{(2)}_{-}(y)+\zeta_-\otimes \eta^{(2)}_{+}(y)\ , \eea where
$\zeta_{+}$ is a generic constant four-dimensional spinor of
positive chirality, while the $\eta^{(a)}_+$ are two particular
six-dimensional spinor fields of positive chirality that
characterise the solution and \bea \zeta_-=(\zeta_+)^*\quad,\quad
\eta^{(a)}_{-}=(\eta^{(a)}_{+})^*\ . \eea In type IIB the two
supersymmetry parameters $\varepsilon_{1,2}$ are MW real spinors
of positive 10d chirality
($\Gamma_{(10)}\varepsilon_{1,2}=\varepsilon_{1,2}$). In this case
\bea \varepsilon_a (y)=\zeta_+\otimes
\eta^{(a)}_{+}(y)+\zeta_-\otimes \eta^{(a)}_{-}(y)\ , \eea where
again $\zeta_-=(\zeta_+)^*$ and
$\eta^{(a)}_{-}=(\eta^{(a)}_{+})^*$. The existence of the internal
spinors $\eta_+^{(1)}$  and $\eta_+^{(2)}$ associated to these
${\cal N}=1$ backgrounds generally specifies an $SU(3)\times
SU(3)$-structure on $T_M\oplus T^\star_M$.

As discussed in \cite{gmpt2}, in order to analyse the supersymmetric conditions for the background, it is convenient to use the bispinor formalism.
Any $O(6,6)$ bispinor $\slashchar{\chi}$ can be written as a sum of antisymmetric products of gamma matrices
\bea
\slashchar{\chi}=\sum_k \frac{1}{k!}\chi^{(k)}_{m_1\ldots m_k}\hat\gamma^{m_1\ldots m_k}\ ,
\eea
which, via the Clifford map, is in  one-to-one correspondence with the formal sum of forms of different degree
\bea
\chi=\sum_k \frac{1}{k!}\chi^{(k)}_{m_1\ldots m_k} dx^{m_1}\wedge\ldots\wedge dx^{m_k}\ .
\eea
We can then associate two pure spinors to our internal spinors $\eta_+^{(1)}$  and $\eta_+^{(2)}$
\bea
\slashchar{\Psi}^+=\eta^{(1)}_+\otimes\eta^{(2)\dagger}_{+}\quad,\quad \slashchar{\Psi}^-=\eta^{(1)}_+\otimes\eta^{(2)\dagger}_{-}
\eea
corresponding to sums of forms of definite parity
\bea
\Psi^+=\sum_{k\geq 0}\Psi^+_{(2k)}\quad\quad \Psi^-=\sum_{k\geq 0}\Psi^-_{(2k+1)}
\eea
Following \cite{gmpt2}, we also define
\bea
||\eta^{(1)}||^2=|a|^2 \quad,\quad ||\eta^{(2)}||^2=|b|^2\ .
\eea

Using the Clifford map, it is  possible to use the gravitino and dilatino Killing conditions to compute
$d\Psi^\pm$ \cite{gmpt2}. The resulting equations are \bea\label{backsusy} e^{-2A+\Phi}(d+H\wedge)\big[
e^{2A-\Phi}\Psi_1\big]&=&dA\wedge \bar \Psi_1 +\frac{e^\Phi}{16}\big[(|a|^2-|b|^2)\hat F +i(|a|^2+|b|^2)\tilde F
\big]\ ,\cr (d+H\wedge)\big[e^{2A-\Phi}\Psi_2\big]&=&0\ , \eea where for type IIA we have \bea\label{defA}
&&\Psi_1=\Psi^-\quad,\quad \Psi_2=\Psi^+\quad {\rm and}\quad F=F_A=F_{(0)}+F_{(2)}+F_{(4)}+F_{(6)}\ , \eea while for type IIB \bea\label{defB} &&\Psi_1=\Psi^+\quad, \quad \Psi_2=\Psi^-\quad {\rm and}\quad
F=F_B=F_{(1)}+F_{(3)}+F_{(5)}\ . \eea  Note that, taking into account the different conventions, the first of
(\ref{backsusy}) has some sign differences with equations (3.2) and (3.3) of \cite{gmpt2}\footnote{\emph{Note added}: We thank the authors of \cite{gmpt2} for private communications confirming the sign mistakes appearing in equations (3.2) and (3.3) in the original version of their paper.}. For this reason we
give some details of the computations leading to (\ref{backsusy}) in appendix \ref{bsusy}. The second condition
means that the generalised  almost complex structure associated to $\Psi_2$ is integrable while in the first
condition the RR fields represent an obstruction to the integrability of the generalised  almost complex
structure associated to $\Psi_1$. Using the gravitino Killing equations one can furthermore show that
\bea\label{backsusy2} d|a|^2=|b|^2dA\quad,\quad d|b|^2=|a|^2dA\ . \eea
As discussed in \cite{gmpt2}, it can be proven that equations (\ref{backsusy}) and (\ref{backsusy2}) are completely equivalent to the full set of supersymmetric Killing conditions and then can be considered as necessary and sufficient conditions to have a supersymmetric background. Furthermore, one has to bear in mind that these equations only make sense if not all of the RR field strengths are vanishing and that in order to have a complete supergravity solution one has to supplement these conditions with the Bianchi identities and the equations of motion for the fluxes \cite{lust}.

The supersymmetry conditions (\ref{backsusy}) and
(\ref{backsusy2}) are identical in form  for type IIA and IIB  and
the two cases are exactly related by the  exchange \bea\label{mir}
\Psi^+\leftrightarrow \Psi^- \quad{\rm and}\quad
F_A\leftrightarrow F_B\ . \eea This relation can be seen as a
generalised mirror symmetry for type II backgrounds with
$SU(3)\times SU(3)$ structure and, as we will see, the conditions
for having supersymmetric branes respect this symmetry, giving
further evidence for it. For further discussions on generalised
mirror symmetry, see e.g. \cite{Fidanza,Ben-Bassat,Jeschek,Grange,Tomasiello,Grana}.

Let us finally remember that these backgrounds contain as subcases
the $SU(3)$ and $SU(2)$ structure backgrounds. In the $SU(3)$ case
we have to require the two $\eta^{(a)}_{+}$  to be linearly
dependent, i.e. $\eta^{(1)}_{+}=a \eta_+$ and
$\eta^{(2)}_{+}=b\eta_+$ for a given six-dimensional spinor field
$\eta_+$, with $\eta_+^{\dagger}\eta_+ = 1$. On the other hand we
have $SU(2)$-structure when  $\eta^{(1)}_{+}$ and
$\eta^{(2)}_{+}$ are never parallel. We refer the reader to the
detailed discussion of these cases given in \cite{gmpt2}.

%%%%%%%%%%%%%%%%%%%%%%%%%%%%%%%%%%%%%%%%%%%%%%%%%%

\section{Supersymmetric D-branes on ${\cal N}=1$ vacua}\label{sec3}
 In general a Dp-brane configuration is  defined by the embedding $\sigma^\alpha\mapsto (x^\mu(\sigma), y^m(\sigma))$, $\alpha=0,\ldots,p$ and preserves a given supersymmetry $\varepsilon$ of the background if it satisfies the condition
\bea\label{susycond} \bar\varepsilon\Gamma_{Dp}=\bar\varepsilon\ ,
\eea where $\Gamma_{Dp}$ is the worldvolume chiral operator
entering the $\kappa$-symmetry transformations \cite{ced,bt}. It
is convenient to use a double spinor convention for both type IIB
and type IIA  where in this last case the two spinors of opposite
chirality are organised in a  two component vector. Using the
explicit form of the $\kappa$-operators in this
notation\footnote{Here we use the $\kappa$-symmetry operators
constructed from T-duality in \cite{mms}, which are identical to
those given in \cite{bt} up to some different overall signs. Their
explicit form in double spinor notation in both IIA and IIB can be
found in \cite{mrvv}.}, the supersymmetry condition reduces to
\bea \hat\Gamma_{Dp}\varepsilon_2=\varepsilon_1\ , \eea where
\bea\label{offdiag} \hat\Gamma_{Dp}=\frac{1}{\sqrt{-\det
(P[G]+{\cal
F})}}\sum_{2l+s=p+1}\frac{\epsilon^{\alpha_1\ldots\alpha_{2l}\beta_{1}\ldots\beta_{s}}}{l!s!2^l}{\cal
F}_{\alpha_1\alpha_2}\cdots{\cal
F}_{\alpha_{2l-1}\alpha_{2l}}\Gamma_{\beta_1\ldots\beta_{s}} \eea
and $\hat\Gamma_{Dp}^{-1}({\cal F})=(-)^{{\rm
Int}[\frac{p+3}{2}]}\hat\Gamma_{Dp}({-\cal F})$. Let us start by
restricting our attention to Dp-branes filling the time plus $q$
flat directions (with no worldvolume flux in these directions),
and wrapping an internal $(p-q)$-cycle. We can then decompose the
above operators into four- and six-dimensional components as
follows \bea\label{splitting}
\hat\Gamma_{Dp}&=&\gamma_{\ul{0\ldots q}}\gamma_{(4)}^{p-q}\otimes
{\hat\gamma}^\prime_{(p-q)}\ ,
%\hat\Gamma_{D(2n)}&=&\gamma_{\ul{0\ldots q}}\gamma_{(4)}^q\otimes {\hat\gamma}^\prime_{(2n-q)}\ ,\cr
%\hat\Gamma_{D(2n+1)}&=&\gamma_{\ul{0\ldots q}}\gamma_{(4)}^{q+1}\otimes \hat{\gamma}^\prime_{(2n+1-q)}\ ,
\eea where \bea \label{ghp}\hat{\gamma}^\prime_{(r)}=
\frac{1}{\sqrt{\det(P[g]+{\cal
F})}}\sum_{2l+s=r}\frac{\epsilon^{\alpha_1\ldots\alpha_{2l}\beta_{1}\ldots\beta_{s}}}{l!s!2^l}{\cal
F}_{\alpha_1\alpha_2}\cdots{\cal F}_{\alpha_{2l-1}\alpha_{2l}}
\hat\gamma_{\beta_1\ldots\beta_{s}}\ , \eea is a unitary operator
acting on the internal spinors.
%Note that
%\bea
%\hat{\gamma}^{\prime -1}_{(r)}({\cal F})=\hat{\gamma}^{\prime \dagger}_{(r)}({\cal F})=(-)^{r(r-1)/2}\hat{\gamma}^\prime_{(r)}(-{\cal F})\ .
%\eea

By considering general Dp-branes in both type IIA/IIB backgrounds
and  using (\ref{spinors}), (\ref{offdiag}) and (\ref{splitting}),
it is possible to see that the supersymmetric condition
(\ref{susycond}) can be split  into the four-dimensional condition
\bea\label{susycond0} \gamma_{\ul{0\ldots q}}\zeta_+=\alpha^{-1}
\zeta_{(-)^{q+1}}\ , \eea and the internal six-dimensional one
\bea\label{susycond1.5}
\hat{\gamma}^\prime_{(p-q)}\eta_{(-)^{p+1}}^{(2)}=\alpha
\eta_{(-)^{q+1}}^{(1)}\ . \eea By consistency with the complex
conjugate of these expressions and the fact that
$\gamma_{\ul{0\ldots q}}^2=-(-)^{\frac{q(q+1)}{2}}$, it can be
seen  that the case $q=0$, i.e. the case where we have an
effective four-dimensional particle, can never be supersymmetric, while for
$q=1,2,3$ one has the condition that $\alpha=e^{i\theta}$, i.e.
$\alpha$ is a pure phase. More explicitly $\theta=0\ {\rm or}\
\pi$ for $q=1$ (effective string), $\theta$ is arbitrary for $q=2$
(domain-wall) and $\theta=-\pi/2$ for $q=3$ (space-time filling
branes). From the unitarity of the operator
$\hat{\gamma}^\prime_{(r)}$, it also follows that we must have the
following constraints on the internal spinors \bea\label{normcond}
||\eta^{(1)}||^2=||\eta^{(2)}||^2\ , \eea and from
(\ref{backsusy2}) we then see that once the condition
(\ref{normcond}) is fulfilled at one point for our backgrounds ,
it is automatically valid at all points.

For the purposes of this work we are interested in spacetime filling branes and hence from this point on we shall consider only these cases. Supersymmetry conditions for the other cases listed above are easily found by reinstating $\theta$-dependence in the appropriate way. In the case of four-dimensional space-time filling branes, the four-dimensional condition is automatically satisfied once we set $\theta=-\pi/2$ and one is left with the following internal conditions
\bea\label{susycond2}
\left\{ \begin{array}{l}
 i\hat{\gamma}^\prime_{(2k)}\eta^{(2)}_+ =\eta^{(1)}_+ \quad\quad,\quad {\rm in\ IIB}\ ,\\
 i\hat{\gamma}^\prime_{(2k+1)}\eta^{(2)}_+ =\eta^{(1)}_- \quad\ ,\quad {\rm in\ IIA}\ . \end{array}\right.
\eea

We would like now to write the supersymmetry conditions
(\ref{susycond2}) in terms of the geometrical objects $\Psi^+$ and
$\Psi^-$ introduced in section \ref{sec1}. In order to do this, it
is useful to decompose the spinorial quantities entering
(\ref{susycond2}) in the basis defined  by \bea\label{base}
\eta_+^{(1)}\quad,\quad \eta_-^{(1)}\quad,\quad
\hat\gamma_m\eta_+^{(1)}\quad {\rm and}\quad
\hat\gamma_m\eta_-^{(1)}\ . \eea By  decomposing the supersymmetry
conditions (\ref{susycond2}) in this basis, one obtains a set of
equations  written in a more geometric fashion in terms of the
pull-back to the worldvolume of $\Psi^+$ and $\Psi^-$. Explicitly,
for even $2k$-cycles we have the conditions \bea\label{gensusy1}
&& \Big\{P[\Psi^+]\wedge
e^{\calf}\Big\}_{(2k)}=\frac{i|a|^2}{8}\sqrt{\det (P[g]+{\cal
F})}d\sigma^1\wedge\ldots\wedge d\sigma^{2k}\ ,\cr
&&\Big\{P[dx^m\wedge \Psi^-+g^{mn}\imath_n\Psi^-]\wedge e^\calf
\Big\}_{(2k)}=0\ , \eea while for odd $(2k+1)$-cycles we have
\bea\label{gensusy2} && \Big\{ P[\Psi^-]\wedge
e^{\calf}\Big\}_{(2k+1)}=\frac{i|a|^2}{8}\sqrt{\det (P[g]+{\cal
F})}d\sigma^1\wedge\ldots\wedge d\sigma^{2k+1}\ ,\cr &&\Big\{
P[dx^m \wedge\Psi^++g^{mn}\imath_n\Psi^+]\wedge
e^\calf\Big\}_{(2k+1)}=0\ . \eea Note that these equations are
identical if we interchange \bea \Psi^+\leftrightarrow \Psi^-\ .
\eea They then respect the generalised mirror symmetry (\ref{mir})
that relates the type IIA and IIB ${\cal N}=1$ supersymmetric
backgrounds we are considering.

In the following section we will discuss the geometrical
interpretation of the supersymmetric conditions (\ref{gensusy1})
and (\ref{gensusy2}). As a preliminary step, it is useful to
observe that they are not independent. Indeed, we obtained these
conditions by expanding (\ref{susycond2}) in the basis
(\ref{base}) and then, using the unitarity of
$\hat\gamma^\prime(\calf)$, it is easy to see that the first
equations of (\ref{gensusy1}) and (\ref{gensusy2}) imply the
seconds. Viceversa, the second conditions determine the first up
to an overall arbitrary (in general, point dependent) phase.
Moreover, once again using the unitarity of
$\hat\gamma^\prime(\calf)$, the first conditions can be
furthermore restricted in such a way that we can characterise the
supersymmetry cycles in the following way: \bea\label{gens1} &&
\Big\{\Im\big(iP[\Psi^+]\big)\wedge e^{\calf}\Big\}_{(2k)}=0\ ,\cr
&&\Big\{P[dx^m\wedge \Psi^-+g^{mn}\imath_n\Psi^-]\wedge
e^\calf\Big\}_{(2k)}=0\ , \eea for even $2k$-cycles, while
\bea\label{gens2} &&\Big\{\Im\big(iP[\Psi^-]\big)\wedge
e^{\calf}\Big\}_{(2k+1)}=0\ ,\cr &&\Big\{P[dx^m
\wedge\Psi^++g^{mn}\imath_n\Psi^+]\wedge e^\calf\Big\}_{(2k+1)}=0\
, \eea for odd $(2k+1)$-cycles.
 Note that these conditions do not strictly speaking imply
that the wrapping brane is supersymmetric but in general it is
supersymmetric only for one orientation. If the RR fields were
turned off, the orientation would be arbitrary because a change of
orientation would amount in considering an anti D-brane instead of
a D-brane or viceversa, and these feel the background fields in
the same way. However, we are considering the case with nontrivial
RR fields. D-branes  and anti D-branes then react to the
background in a different way and the orientation cannot be
ignored, meaning that the conditions given in (\ref{gens1}) and
(\ref{gens2}) are in fact necessary and sufficient only for the
brane to admit at least an orientation making it supersymmetric.

The above conditions can be substituted by the following single
condition that encodes also the necessary orientation requirement:
\bea \label{cc1}&& \Big\{\Re\big(-iP[\Psi^+]\big)\wedge
e^{\calf}\Big\}_{(2k)}=\frac{|a|^2}{8}\sqrt{\det(P[g]+\calf)}d\sigma^1\wedge\ldots\wedge
d\sigma^{2k}\ , \eea for even $2k$-cycles, while for odd
$(2k+1)$-cycles\bea\label{cc2}
&&\Big\{\Re\big(-iP[\Psi^-]\big)\wedge
e^{\calf}\Big\}_{(2k+1)}=\frac{|a|^2}{8}\sqrt{\det(P[g]+\calf)}d\sigma^1\wedge\ldots\wedge
d\sigma^{2k+1}\ . \eea Note that since we are assuming that the
internal spinors have the same norm, in the above expressions we
can write $|a|^2$ in terms of any of the two  pure spinors as
follows \bea
|a|^4=||\Psi||^2=\tr(\slashchar\Psi\slashchar\Psi^\dagger)=8\sum_k|\Psi_{(k)}|^2\
. \eea  We will see in section \ref{gencalsec} that  we can
interpret the equations (\ref{cc1}) and (\ref{cc2}), and then also
(\ref{gens1}) and (\ref{gens2}) plus an appropriate choice of the
orientation, as generalised calibration conditions.

\section{The geometry of the supersymmetric D-branes}\label{geom}

We shall now discuss the geometrical meaning of the second
conditions of (\ref{gens1})  and (\ref{gens2}). As we will see,
supersymmetric  branes wrapping even cycles in type IIB and odd
cycles in type IIA must correspond to a correctly generalised
definition of holomorphic  and  coisotropic branes respectively.
For the cases we are interested in we can adapt the discussion
presented in  \cite{koerber,zabzine} for backgrounds with only
nontrivial NS fields. Furthermore we will use some notions of
generalised complex geometry \cite{hitchin,gualtieri} and a
summary of the basic definitions needed here  are provided in,
for example, \cite{gmpt,zabzine,koerber}.

Let us first recall that, for the general $r$-cycle,  the second   conditions of (\ref{gens1}) and (\ref{gens2})
come from the requirement that $\hat\gamma_{(r)}^\prime(\calf)\eta^{(2)}_+$ must be parallel to
$\eta^{(1)}_{(-)^r}$. It is then possible to see \cite{koerber} that this condition is equivalent to
\bea\label{bcond1} J_1|_\Sigma= (-)^r RJ_2R^{-1}|_\Sigma\ , \eea where $J_1$  and $J_2$ are the almost complex
structures associated to the six-dimensional spinors $\eta_+^{(1)}$ and $\eta_+^{(2)}$ respectively (see Appendix
\ref{conventions} for the basic definitions and properties), and the action of the rotation matrix $R$ on
$T_M|_\Sigma=T_\Sigma\oplus {\cal N}_\Sigma$ is as follows.  If $p_{||}$ and $p_{\perp}$ are the projectors on
the tangent and normal bundle of the brane respectively, then $R$ acts as a reflection in the normal directions
($Rp_{\perp}=p_\perp R=-p_\perp$) while the action of $R$ along $T_\Sigma$ is defined by \bea
p_{||}^{T}(g-\calf)p_{||}=p_{||}(g+\calf)p_{||} R\ , \eea where $\calf$ is now naturally thought of as a section of
$\Lambda^2T_M^\star|_\Sigma$ such that $p_\perp^T \calf =\calf p_\perp=0$. The pure spinors $\Psi^+$ and
$\Psi^-$ are associated to generalised almost complex structures ${\cal J}_+$ and ${\cal J}_-$ on $T_M\oplus
T_M^\star$.  One can prove that these can be written in terms of $J_1$ and $J_2$ as follows
\cite{gualtieri,zabzine,koerber}: \bea {\cal J}_\pm=\frac12\left(
\begin{array}{cc} J_1\mp J_2 & (J_1\pm J_2)g^{-1}  \\  g(J_1\pm J_2) & g(J_1\mp J_2)g^{-1}\end{array}\right)\ .
\eea One can then see that (\ref{bcond1}) is equivalent to the following condition for ${\cal J}_{\pm}$
restricted on $T_M\oplus T^\star_M|_\Sigma$ \bea\label{fineq} {\cal J}_{(-)^{r+1}}={\cal R}^{-1}{\cal
J}_{(-)^{r+1}}{\cal R} \eea where ${\cal R}$ acts in the following way  on $T_M\oplus T_M^\star|_\Sigma$ \bea
{\cal R}=\frac12\left(
\begin{array}{cc} r & 0  \\  \calf r+ r^T\calf & -r^T\end{array}\right)\ , \eea with $r=p_{||}-p_\perp$.

The D-brane worldvolume wrapping the internal cycle $\Sigma$,
specified by the couple $(\Sigma,\calf)$ where $\calf$ is such
that $d\calf=P_\Sigma[H]$, can be seen as a generalised
submanifold as defined by Gualtieri in \cite{gualtieri}.
%\footnote{Note that we are using a different notation from
%\cite{gualtieri}, as we characterise the generalised
%submanifold through the pure worldvolume field-strength $f$ and
%not $\calf=f+P[B]$}.
Gualtieri also defines a {\em generalised tangent bundle}
$\tau_\Sigma^{\calf}$ associated to the brane. The key point is
that the elements $X\in T_M\oplus T^\star_M|_\Sigma$ belonging to
$\tau_\Sigma^{\calf}$ can be characterised by the condition
\cite{zabzine} \bea\label{tgcond} {\cal R}X=X\ . \eea The
subsequent step is to remember that, given an (integrable)
generalised complex structure ${\cal J}$ on $M$, Gualtieri defines
a {\em generalised complex submanifold} as a generalised
submanifold $(\Sigma,\calf)$ with generalised tangent bundle
$\tau_\Sigma^{\calf}$ stable under ${\cal J}$.

From (\ref{fineq}) and (\ref{tgcond}) we arrive at the conclusion that   the second conditions in (\ref{gens1})
and (\ref{gens2}) are each equivalent to the requirement that {\em supersymmetric D-branes wrapping even-cycles
in type IIB and odd-cycles in type IIA must be generalised complex submanifolds with respect to the
(integrable) generalised complex structures ${\cal J}_-$ and ${\cal J}_+$ respectively}. These generalised
complex submanifolds can be seen as the most natural generalisation of complex cycles with $\calf$ of kind
$(1,1)$ in type IIB and of coisotropic cycles in type IIA \cite{kapu,gualtieri}.

%%%%%%%%%%%%%%%%%%%%%%%%%%%%%%%%%%%%%%%%%%%%%%%%%%%%%%%%%%%%%%%%%%%%%%%%%%%%%%%%%%%%%%%%%%%%%%%%%%%%%%%%%%%%%%%%%%%
\section{D-branes on $SU(3)$-structure manifolds}

In this section we pause the discussion of general $SU(3)\times
SU(3)$ structure manifolds to comment on the $SU(3)$ structure
subcase. We recall that this is obtained when we can write
$\eta^{(1)}_+=a\eta_+$ and $\eta^{(2)}_+=b\eta_+$
($\eta_+^\dagger\eta_+=1$), remembering that in order to have
supersymmetric branes we have to fulfil the necessary condition
$|a|=|b|$. In this case the pure spinors $\Psi^\pm$ can be defined
in terms of the almost complex structure $J$ and the $(3,0)$-form
$\Omega$ associated to $\eta_+$ as explained in appendix
\ref{conventions}.  Using the Fierz decomposition it is possible
to show that \bea\label{expo} \eta_{\pm}\otimes
\eta_{\pm}^{\dagger}=\frac{1}8 \slashchar{e^{\mp i J}}\quad,\quad
\eta_{+}\otimes \eta_{-}^{\dagger}=-\frac{i}{8}
\slashchar{\Omega}\quad.\quad \eea We immediately see that in the
$SU(3)$-structure case $\Psi^+$ and $\Psi^-$ reduce to \bea
\Psi^+=\frac{a\bar b}8 e^{-iJ}\quad,\quad
\Psi^-=-\frac{iab}8\Omega\quad.\quad \eea Since we must require
that $|a|=|b|$, we can pose \bea \frac{a}{b}\equiv
e^{i\phi}\quad,\quad \frac{a}{b^*}\equiv e^{i\tau}\ , \eea and the
supersymmetry conditions for the wrapped branes  now read
\bea\label{minksusy1}
&&\Big\{\Im\big(ie^{i\phi}P[e^{-iJ}]\big)\wedge e^{{\cal
F}}\Big\}_{(2k)}=0\ ,\cr &&\Big\{P[dx^m
\wedge\Omega+g^{mn}\imath_n\Omega] \wedge e^\calf\Big\}_{(2k)}=0\
, \eea for even $2k$-cycles, and \bea\label{minksusy2}
&&\Big\{\Im\big(e^{i\tau} P[\Omega]\big)\wedge e^{{\cal
F}}\Big\}_{(2k+1)}=0\ ,\cr && \Big\{P[dx^m \wedge
e^{iJ}+g^{mn}\imath_n e^{iJ}]\wedge e^\calf\Big\}_{(2k+1)}=0\ ,
\eea for odd $(2k+1)$-cycles. Again, these conditions really imply
that it is possible to choose an orientation on the D-brane in
order for it to be supersymmetric and generally reversing the
orientation does not preserve supersymmetry. As in the general
case, they can be substituted by the following equivalent
conditions which also provide the necessary requirement on the
orientation \bea\label{cc3}
&&\Big\{\Re\big(-ie^{i\phi}P[e^{-iJ}]\big)\wedge e^{{\cal
F}}\Big\}_{(2k)}=
\sqrt{\det(P[g]+\calf)}d\sigma^1\wedge\ldots\wedge d\sigma^{2k}\ ,
\eea for even $2k$-cycles, and \bea\label{cc4}
&&\Big\{\Re\big(-e^{i\tau} P[\Omega]\big)\wedge e^{{\cal
F}}\Big\}_{(2k+1)}=\sqrt{\det(P[g]+\calf)}d\sigma^1\wedge\ldots\wedge
d\sigma^{2k+1}\ , \eea for odd $(2k+1)$-cycles. Note that in the
$SU(3)$-case type IIB and IIA backgrounds have complex and
symplectic internal manifolds respectively. The above conditions
have the same form as those derived in \cite{mmms} for branes with
nontrivial worldvolume fluxes on spaces with no fluxes, and can be
seen as their natural generalisation (see also the  discussion in
\cite{marchesano} for the type IIB case). In particular, from the
discussion of the previous section, the second conditions in
(\ref{minksusy1}) and (\ref{minksusy2}) now require that
supersymmetric branes are complex branes with $(1,1)$ field
strength $\calf$ in type IIB and coisotropic branes of the kind
discussed in \cite{kapu} in type IIA (see section 7.2 of
\cite{gualtieri}). Also, the above conditions are obviously
exchanged by the generalised mirror symmetry, that in this case
takes the form \bea e^{i\phi}e^{-iJ}\leftrightarrow -
ie^{i\tau}\Omega\ . \eea

%%%%%%%%%%%%%%%%%%%%%%%%%%%%%%%%%%%%%%%%%
\section{Generalised calibrations for ${\cal N}=1$ vacua}\label{gencalsec}

We shall now proceed to discuss the meaning of the supersymmetry
conditions in the general $SU(3)\times SU(3)$ case. We will see
how the conditions in the form (\ref{cc1}) and (\ref{cc2}) can be
interpreted as generalised calibration conditions.  Then the first
of each pair of conditions (\ref{gens1}) and (\ref{gens2}) encodes
the necessary requirement related to the stability of the
supersymmetric D-brane that must be added to the geometrical
characterisation given in section \ref{geom}.

Let us first of all introduce the appropriate definition of
generalised calibration for the general class of ${\cal N}=1$
manifolds we are considering, starting from the supersymmetry
conditions for four-dimensional space-time filling branes derived
in the previous sections. We will see how it is possible to
naturally introduce a generalised calibration that minimises the
energy and with respect to which supersymmetric cycles are
calibrated. The notion of generalised calibration was first
introduced in \cite{guto} to describe supersymmetric branes on
backgrounds with fluxes, and studied in several subsequent papers
(see for example \cite{marte,ura}). The idea is that the
calibration should minimise the brane energy which does not
necessarily coincide with the volume wrapped by the brane. It has
been shown  in \cite{koerber} how, in the case of pure NS
supersymmetric backgrounds, it is possible to introduce another
notion of generalised  calibration which naturally takes into
account the role of the worldvolume field strength $f$. We will
now see how an analogous definition of generalised calibration can
also be used for general ${\cal N}=1$ backgrounds with nontrivial
RR fluxes.

We define a {\em generalised  calibration}  as a sum of forms of
different degree $\omega=\sum_k\omega_{(k)}$ such that
$d_H\omega=(d+H\wedge)\omega=0$ and \bea\label{gencal}
P_\Sigma[\omega]\wedge e^\calf \leq {\cal E}(\Sigma,\calf)\ , \eea
for any D-brane $(\Sigma,\calf)$ characterised by the wrapped
cycle $\Sigma$ and the worldvolume field strength $\calf$ and with
energy density ${\cal E}$ {}\footnote{This definition is
completely equivalent to the definition used in \cite{koerber}
where a generalised calibration $\tilde\omega$ is closed, i.e.
$d\tilde\omega=0$, and satisfies the relation
$P_\Sigma[\omega]\wedge e^f \leq {\cal E}(\Sigma,\calf)$. The two
generalised calibrations are obviously related by
$\tilde\omega=\omega\wedge e^B$. We prefer our choice as it
involves the worldvolume field-strength $f$ only through the gauge
invariant combination $\calf$.}. In (\ref{gencal}) and all other
expressions in this section involving sums of forms of different
degree on the cycle wrapped by the brane, we understand that only forms of rank equal to the
dimension of the cycle are selected.
Furthermore, the inequalities between these forms refer to the
associated scalar components in the one-dimensional base given by
the standard (oriented) volume form.

A D-brane $(\Sigma,\calf)$ is then {\em calibrated in a
generalised sense} by  $\omega=\sum_k \omega_{(k)}$, if it
satisfies the condition \bea P_\Sigma[\omega]\wedge e^\calf={\cal
E}(\Sigma,\calf)\ . \eea Since the generalised  calibration
$\omega$ is $d_H$-closed, one can immediately prove that the
saturation of the calibration bound is a minimal energy condition.
Let $E(\Sigma,\calf)$ be the four-dimensional energy density of a
calibrated wrapped D-brane $(\Sigma,\calf)$. Consider a continuous
deformation to a different brane configuration
$(\Sigma^\prime,\calf^\prime)$ such that we can take a chain
${\cal B}$ and a field-strength $\hat\calf$ on it (with
$d\hat\calf=P_{\cal B}[H]$), such that $\partial {\cal
B}=\Sigma-\Sigma^\prime$ and the restriction of $\hat\calf$ to
$\Sigma$ and $\Sigma^{\prime}$ gives $\calf$ and $\calf^{\prime}$
respectively. We then have \bea E(\Sigma,\calf)&=&\int{\cal
E}(\Sigma,\calf)=\int_\Sigma P[\omega]\wedge e^\calf = \cr &=&
\int_{\cal B}P[d_H\omega]\wedge e^{\hat\calf}
+\int_{\Sigma^\prime} P[\omega]\wedge e^{\calf^\prime}=\cr &=&
\int_{\Sigma^\prime} P[\omega]\wedge e^{\calf^\prime}\leq
\int{\cal E}(\Sigma^\prime,\calf^\prime) =
E(\Sigma^\prime,\calf^\prime)\ . \eea A calibration condition can
then be seen as a stability condition for a D-brane under
continuous deformations.

We will now see how the supersymmetry conditions in (\ref{cc1}) and (\ref{cc2}) can be rephrased as generalised  calibration conditions. In order to prove this, we have to construct the generalised  calibration appropriate  to our case. Let us start by recalling that we are restricting to the case in which $\eta^{(1)}$ and $\eta^{(2)}$ have the same norm. Then the standard Schwarz inequality
\bea
||i\hat\gamma^\prime_{(r)}(\calf)\eta^{(2)}_+ + \eta^{(1)}_{(-)^r}||\leq ||i\hat\gamma^\prime_{(r)}(\calf)\eta^{(2)}_+||
+||\eta^{(1)}_{(-)^r}||\ ,
\eea
implies that we have the following completely general inequalities
 \bea
\Re\big[i\eta^{(1)\dagger}_+\hat\gamma^\prime_{(2k)}(\calf)\eta^{(2)}_{+}\big]\leq
|a|^2\quad,\quad
\Re\big[i\eta^{(1)\dagger}_-\hat\gamma^\prime_{(2k+1)}(\calf)\eta^{(2)}_{+}\big]\leq
|a|^2\ , \eea which, remembering (\ref{susycond2}), are clearly
saturated when we are considering supersymmetric cycles. Using
expression (\ref{ghp}) for $\hat\gamma^\prime_{(r)}$  it is not
difficult to see that from these relations we obtain the
conditions \bea\label{dis1} \Big\{\Re\big(-iP[\Psi^+]\big)\wedge
e^{\calf}\Big\}_{(2k)}&\leq&\frac{|a|^2}{8}\sqrt{\det(P[g]+\calf)}
d\sigma^1\wedge\ldots\wedge d\sigma^{2k}\ ,\cr
\Big\{\Re\big(-iP[\Psi^-]\big)\wedge
e^{\calf}\Big\}_{(2k+1)}&\leq&\frac{|a|^2}{8}\sqrt{\det(P[g]+\calf)}
d\sigma^1\wedge\ldots\wedge d\sigma^{2k+1}\ . \eea Once we impose
that the D-branes must wrap generalised  complex submanifolds in
$M$, one sees that  requiring the inequalities in (\ref{dis1}) to
be saturated is equivalent to requiring that the D-branes we are
considering satisfy the  supersymmetry conditions  (\ref{cc1}) and
(\ref{cc2}).

We would now like to use these inequalities to construct a
generalised calibration for this space-time filling branes. Given
the RR field-strength ansatz specified in (\ref{RRdec}), we can
analogously decompose the RR potentials in the following way
\bea\label{rrdec} C_{(n)}=\hat C_{(n)}+dx^0\wedge\ldots\wedge dx^3
\wedge e^{4A}\tilde C_{(n-4)}\ , \eea and then express the
internal RR field strengths in terms of the internal RR potentials
\bea && \hat F_{(k+1)}=d\hat C_{(k)}+H\wedge \hat C_{(k-2)}\ ,\cr
&& \tilde F_{(k+1)}=d\tilde C_{(k)}+H\wedge \tilde
C_{(k-2)}+4dA\wedge \tilde C_{(k)}\ . \eea Our space-time filling
branes couple only to the ``tilded'' RR fields. Since we are
considering static configurations, we can extract from the
Dirac-Born-Infield plus Chern-Simons action the following
effective energy density for a space-time filling brane wrapping
an internal $n$-cycle \bea\label{endens} {\cal
E}=e^{4A}\Big\{e^{-\Phi}\sqrt{\det(P[g]+\calf)}d\sigma^1\wedge\ldots\wedge
d\sigma^n - \Big(\sum_k P[\tilde C_{(k)}]\wedge
e^\calf\Big)_{(n)}\Big\}\ , \eea where for simplicity we have
omitted the overall factor given by the D-brane tension. We can
now write the inequalities (\ref{dis1}) in terms of a lower bound
on the energy density \bea\label{caldis} && P[\omega]\wedge
e^\calf\leq {\cal E}\ , \eea where we have used the sum of forms
of different degrees $\omega=\sum_k \omega_{(k)}$ given by
\bea\label{cal1} \omega_{IIA}&=&
e^{4A}\Big[\Re\Big(\frac{-8i}{|a|^2}e^{-\Phi}\Psi^-\Big)-\sum_k
\tilde C_{(2k+1)}\Big]\ ,\cr \omega_{IIB}&=&e^{4A}\Big[
\Re\Big(\frac{-8i}{|a|^2}e^{-\Phi}\Psi^+\Big)-\sum_k \tilde
C_{(2k)}\Big]\ . \eea Note that in the left hand side of
(\ref{caldis}) one can completely factorise  the contributions of
the background quantities through the pullback on the cycle of
$\omega$ and $B$, and the contribution from the worldvolume
field-strength $f$.

It is clear from (\ref{caldis}) that the $\omega$'s  defined in
(\ref{cal1}) represent a good candidate for  generalised
calibrations as described at the beginning of this section. To
prove that this is indeed the case, it remains to show that the
$\omega$'s in (\ref{cal1})  are $d_H$-closed. In order to do this,
it will be enough to use the equations (\ref{backsusy}) and
(\ref{backsusy2}), which characterise our ${\cal N}=1$
backgrounds.

Let us impose the vanishing of the $d_H$-differential of the $\omega$'s  defined in (\ref{cal1}). This gives the
following condition to have properly defined calibrations \bea\label{calcond} d_H\omega_{IIA}=0&\Leftrightarrow&
\big[d+(H+4dA)\wedge\big]\Big[\frac{1}{|a|^2} e^{-\Phi}\Re\Big(i\Psi^-\Big)\Big]= -\frac18
\sum_{k=0,1,2,3}\tilde F_{(2k)}\ ,\cr d_H\omega_{IIB}=0&\Leftrightarrow&
\big[d+(H+4dA)\wedge\big]\Big[\frac{1}{|a|^2} e^{-\Phi}\Re\Big(i\Psi^+\Big)\Big]=
-\frac{1}{8}\sum_{k=0,1,2}\tilde F_{(2k+1)}\ .
 \eea One immediately sees that
the conditions (\ref{backsusy}) and (\ref{backsusy2}) imply that
the above requirements are indeed satisfied. This concludes our
proof that our ${\cal N}=1$ backgrounds are generalised complex
manifolds with generalised calibrations  defined in (\ref{cal1}),
such that supersymmetric four-dimensional spacetime filling branes wrap
generalised complex submanifolds, which are also generalised
calibrated.

One can get more intuition on the structure of the above
generalised calibrations  by considering the $SU(3)$-structure
subcase. The generalised calibrations then take the form
\bea\label{calsu3} \omega_{IIA}&=&e^{4A}\Big[
\Re\Big(-e^{i\tau}e^{-\Phi}\Omega\Big)-\sum_k \tilde
C_{(2k+1)}\Big]\ ,\cr \omega_{IIB}&=&e^{4A}\Big[
\Re\Big(-ie^{i\phi}e^{-\Phi}e^{-iJ}\Big)-\sum_k \tilde
C_{(2k)}\Big]\ . \eea  We then explicitly see how these
calibrations generalise the usual calibrations in Calabi-Yau
spaces through crucial modifications introduced by the nontrivial
dilaton, warp-factor and fluxes.

Note also that the generalised calibrations (\ref{cal1}) are
naturally related by the mirror symmetry (\ref{mir}),  if we
exchange $\sum_{k}\tilde C_{(2k)}$ and $\sum_k\tilde C_{(2k+1)}$.
These sums can be seen as H-twisted potentials of the sums of
internal field strengths $\tilde F_{A}$ and $\tilde F_{B}$ as
defined in (\ref{defA}) and (\ref{defB}). If we think in terms of
untwisted quantities we then get a mirror symmetry for the
potentials of the form \bea \sum_{k}\tilde C_{(2k)}\wedge
e^B\leftrightarrow \sum_{k}\tilde C_{(2k+1)}\wedge e^B\ , \eea
which clearly recalls the form of the transformation rules of the
RR-potentials under T-duality.

Let us observe that the generalised calibration $\omega$  defined
above is a sum of forms which are not generally globally defined,
since they are not invariant under the RR gauge transformations.
Indeed, consider the gauge transformation \bea \sum_n \delta\tilde
C_{(n)}=e^{-4A}d_H\lambda\ ,\eea preserving the decomposition
(\ref{rrdec}), where $\lambda$ is a sum of even (odd) forms for
type IIA (IIB). Then $\omega$ transforms as $\omega\rightarrow
\omega -d_H\lambda$, since it is related to the D-brane energy
density which naturally depends on the RR gauge potentials. As an
alternative, we could also introduce an equivalent globally
defined generalised calibration
$\hat\omega=\sum_n\hat\omega_{(n)}$ which is more in the spirit of
that adopted in \cite{guto}. First, in our class of backgrounds,
$\hat\omega$ is no longer $d_H$ closed, but must satisfy the
condition \bea \label{agc}d_H\hat\omega=e^{4A}\sum_k\tilde
F_{(k)}\ .\eea Secondly, the energy density minimisation condition
(\ref{caldis}) is replaced by the condition \bea\label{agc2}
P_\Sigma[\hat\omega]\wedge e^\calf \leq
e^{4A-\Phi}\sqrt{\det(P[g]+\calf)}d\sigma^1\wedge\ldots d\sigma^n\
,\eea for any D-brane $(\Sigma,\calf)$ wrapping an internal
$n$-dimensional cycle. It is clear from our previous discussion
that such an alternative generalized calibration is given by
\bea\label{altgencal} \hat\omega
=-\frac{8e^{4A-\Phi}}{|a|^2}\Re(i\Psi)\ ,\eea where $\Psi=\Psi^+$
for type IIB and $\Psi=\Psi^-$ for type IIA. We obviously have
that $\omega=\hat\omega-e^{4A}\sum_n\tilde C_{(n)}$ and the
alternative generalized calibration $\hat\omega$ can be
essentially identified with the imaginary part of the
non-integrable pure spinor characterising the ${\cal N}=1$
background considered. 

As we are assuming $|a|=|b|$, 
the condition (\ref{agc})
is equivalent to the imaginary part of the first of the background 
supersymmetry conditions (\ref{backsusy}), thus giving a physical 
interpretation for it. It is nice to note that an analogous conclusion can be reached 
for the remaining equations in (\ref{backsusy}). Indeed, we have 
seen in section \ref{sec3} how we could also consider supersymmetric
branes filling only two or three flat space-time directions, giving rise to
an effective string or domain wall respectively with appropriately chosen phases $\a$ in (\ref{susycond0}). One can then repeat the arguments of this section
for these cases, with the generalised calibrations now be given by
\bea
\omega^{(string)}=\frac{8e^{2A-\Phi}}{|a|^2}\Re(\Psi_1)\quad,\quad 
\omega^{(DW)}=\frac{8e^{3A-\Phi}}{|a|^2}\Re(e^{i\theta}\Psi_2)\ ,
\eea
where $\Psi_1=\Psi^+$ ($\Psi^-$) and $\Psi_2=\Psi^-$ ($\Psi^+$)  
for type IIB (IIA), and  $\theta$ is an arbitrary (constant) phase.    
The generalised calibrations $\omega^{(string)}$ and $\omega^{(DW)}$
now satisfy the condition (\ref{agc2}) with $e^{4A}$ substituted by \
$e^{2A}$ and $e^{3A}$ respectively. Furthermore, they must now be $d_H$-closed, 
since the coupling to the background RR-fields vanishes for these configurations . It is then easy to see that the condition $d_H\omega^{(string)}=0$ is equivalent to the real part of the first of (\ref{backsusy}) (with $|a|=|b|$), while $d_H\omega^{(DW)}=0$ for any $\theta$ is equivalent to the second of (\ref{backsusy}) . We then see how, in the subcase where the two internal spinors have the same norm, the background supersymmetry conditions (\ref{backsusy}) have a physical interpretation as conditions for the existence of generalised calibrations for the allowed supersymmetric D-brane configurations. This correspondence between background supersymmetry conditions and generalized calibrations has been extensively discussed in \cite{marte}
and we see here how it works perfectly in the cases we have considered.

\section{Conclusions}

In this paper we have studied the conditions for having  supersymmetric D-branes in type II backgrounds  with general NS and RR fields preserving four-dimensional Poincar\'e invariance and ${\cal N}=1$ supersymmetry, focusing on D-branes filling the four flat directions. It turns out that
the supersymmetry conditions for D-branes obtained from $\kappa$-symmetry can be elegantly expressed in terms of the two pure spinors that define the $SU(3)\times SU(3)$-structure on the internal six-dimensional manifold. We have shown that the supersymmetry conditions give two important pieces of information on the supersymmetric D-branes, regarding the geometry and the stability of the branes, as happens in absence of fluxes, and involving the two pure spinors separately.

Firstly, the D-brane must wrap a generalised  complex submanifold
defined with respect to the integrable generalised complex
structure of the internal manifold. This can be introduced thanks
to the integrability of one of the two pure spinors coming from
the requirement of ${\cal N}=1$ supersymmetry. The $SU(3)$
structure subcase provides a clear example where this condition
means that the brane must wrap a holomorphic cycle with  $(1,1)$
field strength $\calf$ in type IIB and a coisotropic cycle of the
kind discussed in \cite{kapu,gualtieri} in type IIA. In the more
general $SU(3)\times SU(3)$ case the equivalent type IIA/IIB
identifications become slightly mixed.

Secondly, on the wrapped internal $n$-cycle one must furthermore
impose a condition of the form
$\{\Im(P[i\Psi])\wedge\calf\}_{(n)}=0$, where $\Psi$ is the
non-integrable pure spinor. This condition is related to the
stability of the D-brane. Note that it is the non-integrable pure
spinor that now plays the relevant role and the fact that it
should be connected to some dynamical information for the D-branes
can be linked to the role of the nontrivial RR-fields as
obstructions to the integrability of the pure spinor. Then, a
supersymmetric D-brane configuration must satisfy the above two
conditions, plus an appropriate choice of its orientation which is
in general not arbitrary due to presence of nontrivial background
RR fields. 

The above requirements that characterise supersymmetric D-branes
are equivalent to the condition that the D-brane must be
calibrated in a generalised sense with respect to an appropriate
definition of generalised calibration. This encodes a requirement
of minimisation of the energy of the brane and involves the
non-integrable pure spinor. The non-integrability of this pure
spinor is due to the non-vanishing RR-fields, which also couple to
D-branes and so must enter the associated generalised
calibrations. Then one sees that the non-integrability of the pure
spinor is exactly what is needed to compensate for the presence of
the RR terms in the generalised calibration in order for it to be
well defined. This strict relation between the non-integrable pure
spinor and  a generalised calibration can be made even more
explicit by using the equivalent alternative definition of
generalised calibration given in (\ref{agc}) and (\ref{agc2}).
Furthermore, as we discuss at the end of section \ref{gencalsec}, by 
considering D-branes filling only two or three flat directions, 
the conditions for the existence of well defined calibrations associated to
supersymmetric D-branes are completely equivalent to the background supersymmetry 
conditions (\ref{backsusy}), thus giving a clear physical interpretation for them.

To conclude, it is intriguing to see how the two pure spinors can
be fruitfully used in the description of the geometrical and
stability features of supersymmetric D-branes. Also, all the
results discussed in this paper confirm the interpretation of the
symmetry (\ref{mir}) relating type IIA and IIB backgrounds as a
generalised mirror symmetry, exchanging also odd and even
dimensional supersymmetric cycles and the corresponding
generalised calibrations. These results may hide some deeper
insight into the understanding of string theory on general
backgrounds with fluxes and its relation to generalised
geometry.

%\vskip 1.3cm
\newpage

\begin{center}
{\large  {\em Acknowledgements}}
\end{center}
We would like to thank G. Bonelli, M. Gra{\~n}a, F. Marchesano, J. Sparks, K. Stelle and A. Uranga. We especially thank A. Van Proeyen for his comments and careful reading of the manuscript. This
work is supported in part by the Federal Office for Scientific,
Technical and Cultural Affairs through the "Interuniversity
Attraction Poles Programme -- Belgian Science Policy" P5/27 and by
the European Community's Human Potential Programme under contract
MRTN-CT-2004-005104 `Constituents, fundamental forces and
symmetries of the universe'. PS is supported by the Leverhulme
Trust.

%\vskip 2cm

\begin{appendix}
\section{Basic definitions for $SU(3)$ structure manifolds}\label{conventions}
In this section we will review some basic facts about an
$SU(3)$-structure manifold $M$, that is characterised by the
existence of a globally defined spinor $\eta_+$, such that
$||\eta||^2=|a|^2$ (for a nice review on this subject see for
example \cite{glmw}). This spinor allows one to introduce an
associated almost complex structures with respect to which the six-dimensional
metric $g_{mn}$ is Hermitian. For our purposes the most useful
choice is given by \bea\label{complex}
J_{mn}=-\frac{i}{|a|^2}\eta_+^{\dagger}\hat\gamma_{mn}\eta_+\ .
\eea Using the Fierz identities, it is possible to show that \bea
J_m{}^p J_p{}^n=-\delta_m^n\quad,\quad J_m{}^p J_n{}^q
g_{pq}=g_{mn}\ . \eea This almost complex structure allows one to
introduce the projector on holomorphic indices \bea {\cal
P}_m{}^n=\frac12(\delta_m^n-iJ_m{}^n)\ , \eea and the associated
anti-holomorphic projector $\bar{\cal P}_m{}^n=({\cal
P}_m{}^n)^*$. One can then split $r$-forms in $(p,q)$-forms, with
$p+q=r$, in the standard way.

The following relations hold \bea
\eta^{\dagger}_+\hat\gamma^m\hat\gamma_n\eta_+=2|a|^2\bar{\cal
P}^m{}_n\quad,\quad
\eta^{\dagger}_-\hat\gamma^m\hat\gamma_n\eta_-=2|a|^2{\cal
P}^m{}_n\ . \eea Then $\hat\gamma_m\eta_+={\cal
P}_m{}^n\hat\gamma_n\eta_+$ (it is of the kind $(1,0)$ in the
index $m$),  and the base (\ref{base}) is indeed eight
dimensional. The general six-dimensional Dirac spinor $\chi$ can
then be decomposed as \bea\label{decomp} \chi=\lambda_1\eta_+
+\lambda_2\eta_- + \xi^m_1\hat\gamma_m\eta_+
+\xi^m_2\hat\gamma_m\eta_-\ , \eea where $\xi^m_1$ is a
$(1,0)$-vector (${\cal P}_m{}^n\xi^m_1=\xi^n_1$) and  $\xi^m_2$ is
a $(0,1)$-vector ($\bar{\cal P}_m{}^n\xi^m_1=\xi^n_1$). Then,
\bea\label{decomp2}
&&\lambda_1=\frac1{|a|^{2}}\eta_+^{\dagger}\chi\quad ,\quad
\lambda_2=\frac1{|a|^2}\eta_-^{\dagger}\chi\quad ,\cr &&
\xi^m_1=\frac1{2|a|^2}\eta_+^{\dagger}\hat\gamma^m\chi\quad ,\quad
\xi^m_2=\frac1{2|a|^2}\eta_-^{\dagger}\hat\gamma^m\chi\ . \eea

Analogously to the $CY_3$ case, we can also introduce a $(3,0)$ form $\Omega$ defined by
\bea\label{hol3}
\Omega_{mnp}=-\frac{i}{a^2}\eta^{\dagger}_- \hat\gamma_{mnp}\eta_+\ .
\eea
By applying Fierz identities it is possible to see that
\bea\label{condsu3}
\frac{1}{3!} J\wedge J\wedge J=\frac{i}{8}\Omega\wedge \bar\Omega \quad,\quad J\wedge \Omega=0\ ,
\eea
as for Calabi-Yau manifolds. The existence of a globally defined non-degenerate (real) $J$ and a globally defined non-degenerate (complex) $\Omega$ satisfying the conditions (\ref{condsu3}) actually characterises $SU(3)$-structure manifolds. In our case we are considering the more general case of internal  manifolds $M$ with $SU(3)\times SU(3)$-structure group for $T_M\oplus T_M^\star$. This contains as subcases the $SU(3)$-structure manifolds case and the even more restricted manifolds with $SU(2)$-structure, that contain two different independent $SU(3)$ structures and requires the vanishing of the Euler characteristic $\chi$ of $M$.

\section{Background supersymmetry conditions}
\label{bsusy} Here we recall the way to derive the background
supersymmetry conditions as given in (\ref{backsusy}). In
\cite{gmpt2} it was first shown how the Killing equations can be
written in this elegant form in terms of the pure spinors
$\Psi^\pm$. However, repeating the calculation we find a slightly
different result from that given in \cite{gmpt2}. For this reason we present some details of the
calculation leading to (\ref{backsusy}).

We will follow the method described in \cite{gmpt,gmpt2} which
uses the democratic formalism of \cite{bkorp}, and in particular
we adopt their conventions in the following calculation. The
result can be translated back to our convention by using the rules
specified in footnote 2 on page 4. It will be sufficient to
consider only the type IIA case (the type IIB case is completely
analogous), for which the supersymmetry transformations for the
gravitino are \bea \label{gsst} \delta \psi_M = \nabla_M
\varepsilon + \frac{1}{4} H_M \Gamma_{(10)} \varepsilon +
\frac{1}{16} e^{\Phi} \sum_n  \slashchar{F}_{(2n)} \, \Gamma_{M}
\Gamma_{(10)}^n \sigma_1 \, \varepsilon  \, , \eea where
$\varepsilon$ is  defined in (\ref{spinors}),
$\slashchar{\partial}  \equiv \hat{\gamma}^m\partial_m$ and  $H_M
\equiv \frac{1}{2}H_{MNP} \Gamma^{NP}$. The modified RR field
strengths $F_{(2n)}=dC_{(2n-1)} - H \wedge C_{(2n-3)}$ are related
by the conditions $F_{(2n)}  = (-)^n \star_{10} F_{(10-2n)}$.
%\label{sd10}
We will also need  the modified dilatino transformation \bea
\Gamma^M \delta \psi_M - \delta \lambda = \Big(\slashchar{\nabla}
- \slashchar{\partial} \Phi  + \frac14 \slashchar{H} \Gamma_{(10)}
\Big) \, \varepsilon \, . \label{md} \eea Using the 4d+6d
decomposition of spinors, field strengths and gamma matrices
described in section 2, we may rewrite the transformations above
as conditions upon the internal spinors $\eta^{(a)}_\pm$. For the
external component of the gravitino transformation $\delta
\psi_{\mu}$ one then finds \bea
\frac{1}{2} \slashchar{\partial} A \eta^{(1)}_+ + \frac{e^{\Phi}}{16}\left(\hat{\slashchar{F}}_{A1} + i\tilde{\slashchar{F}}_{A1} \right) \eta^{(2)}_- ~=~ 0 \, , \\
\frac{1}{2} \slashchar{\partial} A \eta^{(2)}_- + \frac{e^{\Phi}}{16}\left(\hat{\slashchar{F}}_{A2} + i\tilde{\slashchar{F}}_{A2} \right) \eta^{(1)}_+ ~=~ 0 \, ,
\eea
where $F_{A1} = F_{(0)} - F_{(2)} + F_{(4)} - F_{(6)}$ and $F_{A2} = F_{(0)} + F_{(2)} + F_{(4)} + F_{(6)}$. The corresponding decomposition on the internal component $\delta \psi_m$ gives
\bea
\nabla_m \eta^{(1)}_+ + \frac{1}{4} H_m \eta^{(1)}_+ + \frac{e^{\Phi}}{16}\left(\hat{\slashchar{F}}_{A1} - i\tilde{\slashchar{F}}_{A1} \right) \hat{\gamma}_m \eta^{(2)}_- ~=~ 0 \, , \\
\nabla_m \eta^{(2)}_- - \frac{1}{4} H_m \eta^{(2)}_- + \frac{e^{\Phi}}{16}\left(\hat{\slashchar{F}}_{A2} - i\tilde{\slashchar{F}}_{A2} \right) \hat{\gamma}_m \eta^{(1)}_+ ~=~ 0 \, ,
\eea
and for the modified dilatino transformation we find
\bea
\left( \slashchar{\nabla} +\frac{1}{4}\slashchar{H} + 2\slashchar{\partial}A - \slashchar{\partial}\Phi \right) \eta^{(1)}_+ ~=~ 0 \, , \\
\left( \slashchar{\nabla} -\frac{1}{4}\slashchar{H} +
2\slashchar{\partial}A - \slashchar{\partial}\Phi \right)
\eta^{(2)}_+ ~=~ 0 \, . \eea Consider the exterior derivative of
the Clifford(6,6) spinors $\Psi^{\pm}$ in terms of bispinors,
given by \bea\label{c66} d\Psi^{\pm}= dx^m \wedge \nabla_m
\Psi^{\pm}= dx^m \wedge \left[ (\nabla_m \eta^{(1)}_{+}) \otimes
\eta_{\pm}^{(2) \dagger} + \eta^{(1)}_{+} \otimes (\nabla_m
\eta_{\pm}^{(2)})^\dagger \right] \, . \eea We shall concentrate on
$d\Psi^-$, with the aim of deriving the first expression in
(\ref{backsusy}) for type IIA. Using the definition of
Clifford(6,6) spinor representations given in \cite{gmpt,gmpt2},
we can rewrite (\ref{c66}) as \bea\label{c662} 2d\Psi^- =
\slashchar{\nabla} \eta^{(1)}_+ \otimes
{\eta^{(2)\dagger}_-} -  \eta^{(1)}_+ \otimes
(\slashchar{\nabla}{\eta^{(2)}_-})^{\dagger} +
\hat{\gamma}^m\eta^{(1)}_+ \otimes
(\nabla_m{\eta^{(2)}_-})^{\dagger} - \nabla_m\eta^{(1)}_+ \otimes
{\eta^{(2)\dagger}_-}\hat{\gamma}^m \ . \eea One can now
use the decomposition of internal gravitino and modified dilatino
supersymmetry transformations given above to evaluate the
right-hand side.  We now form the following bispinors from the
external gravitino transformation \bea
\frac{1}{2} \slashchar{\partial} A \eta^{(1)}_- \otimes {\eta^{(2)\dagger}_+} - \frac{e^{\Phi}}{16}\left(\hat{\slashchar{F}}_{A1} - i\tilde{\slashchar{F}}_{A1} \right) \eta^{(2)}_+ \otimes {\eta^{(2)\dagger}_+} ~=~ 0 \, , \\
\frac{1}{2} \eta^{(1)}_- \otimes {\eta^{(2)\dagger}_+}
\slashchar{\partial} A - \frac{e^{\Phi}}{16}\eta^{(1)}_- \otimes
{\eta^{(1)\dagger}_-} \left(\hat{\slashchar{F}}_{A1} +
i\tilde{\slashchar{F}}_{A1} \right) ~=~ 0 \, . \eea These two
quantities can be added together with (\ref{c662}) to give the
following expression, \bea 2d\Psi^- &=&
\left(\slashchar{\partial}\Phi -\frac{1}{4}\slashchar{H} -
2\slashchar{\partial}A \right) \eta^{(1)}_+ \otimes
{\eta^{(2)\dagger}_-} - \eta^{(1)}_+ \otimes
{\eta^{(2)\dagger}_-}\left(\slashchar{\partial}\Phi
-\frac{1}{4}\slashchar{H} - 2\slashchar{\partial}A \right)
\nonumber \\ &-& \frac{1}{4}\hat{\gamma}^m\eta^{(1)}_+ \otimes
{\eta^{(2)\dagger}_-}H_m + \frac{1}{4}H_m\eta^{(1)}_+ \otimes
{\eta^{(2)\dagger}_-}\hat{\gamma}^m -
\slashchar{\partial}A\eta^{(1)}_- \otimes {\eta^{(2)\dagger}_+} +
\eta^{(1)}_- \otimes {\eta^{(2)\dagger}_+}\slashchar{\partial}A
\nonumber \\ &-&
\frac{e^{\Phi}}{16}\left(\hat{\gamma}^m\eta^{(1)}_+ \otimes
{\eta^{(1)\dagger}_+}\hat{\gamma}_m + 2\eta^{(1)}_- \otimes
{\eta^{(1)\dagger}_-}\right)(\hat{\slashchar{F}}_{A1} +
i\tilde{\slashchar{F}}_{A1}) \nonumber \\ &+&
\frac{e^{\Phi}}{16}(\hat{\slashchar{F}}_{A1} -
i\tilde{\slashchar{F}}_{A1})\left(\hat{\gamma}^m\eta^{(2)}_-
\otimes {\eta^{(2)\dagger}_-}\hat{\gamma}_m + 2\eta^{(2)}_+
\otimes {\eta^{(2)\dagger}_+}\right)\ . \eea Making  some
manipulations, using the fact that
$\hat\gamma_{(6)}\hat{\slashchar{F}}_{A1}=-i\tilde{\slashchar{F}}_{A1}$
and going back to the notations used in this paper as specified in
the footnote 2, the Clifford map allows one to  write this equation in
the form  (\ref{backsusy}).

\end{appendix}

%%%%%%%%%%%%%%%%%%%%%%%%%%%%%%%%%%%%%%%%%%%%%%

%%%%%%%%%%%%%%%%%%%%%%%%%%%%%%%%%%%%%%%%%%%%%%%%%

\end{document}